\documentclass{svproc}
\usepackage{url}

\usepackage{cite}
\usepackage{enumitem}
\usepackage{amsmath,amssymb,amsfonts}
\usepackage{algorithmic}
\usepackage{wrapfig}
\usepackage{graphicx}
\usepackage{caption, subcaption}
\usepackage{array, booktabs, multirow}
\usepackage{textcomp}
\usepackage[usenames,dvipsnames,svgnames,table]{xcolor}
\definecolor{low}{rgb}{0.67, 0.88, 0.69}
\definecolor{medium}{rgb}{1.0, 0.99, 0.82}
\definecolor{high}{rgb}{0.99, 0.56, 0.67}

\begin{document}
\mainmatter              
\title{A GDPR-compliant Risk Management Approach based on Threat Modelling and ISO 27005}
\titlerunning{A GDPR-Compliant Threat-Model-Based Risk Management Approach}  
%
\author{Denys A. Flores\textsuperscript{*} and Ricardo Perugachi}
\authorrunning{Denys A. Flores and Ricardo Perugachi} 
%
\tocauthor{Denys A. Flores and Ricardo Perugachi}
\institute{Departamento de Inform\'atica y Ciencias de la Computaci\'on (DICC), Escuela Polit\'ecnica Nacional, Quito, Ecuador\\
\email{\{denys.flores,ricardo.perugachi\}@epn.edu.ec},\\ WWW home page\textsuperscript{*}:
\texttt{https://www.linkedin.com/in/denysflores/}
}
\maketitle              

\begin{abstract}
Computer systems process, store and transfer sensitive information which makes them a valuable asset. Despite the existence of standards such as ISO 27005 for managing information risk, cyber threats are increasing, exposing such systems to security breaches, and at the same time, compromising users’ privacy. However, threat modelling has also emerged as an alternative to identify and analyze them, reducing the attack landscape by discarding low-risk attack vectors, and mitigating high-risk ones. In this work, we introduce a novel threat-modelling-based approach for risk management, using ISO 27005 as a baseline for integrating ISO 27001/27002 security controls with privacy regulations outlined in the European General Data Protection Regulation (GDPR). In our proposal, risk estimation and mitigation is enhanced by combining STRIDE and attack trees as a threat modelling strategy. Our approach is applied to an IoT case study, where different attacks are analyzed to determine their risk levels and potential countermeasures.
\end{abstract}
\section{Introduction}
Currently, computer systems are an important asset for ensuring the success of any organization. As such, securing the information they process, store and transfer is a paramount objective for guaranteeing a steady increase in the market \cite{iso2}. These systems will always be challenged by emerging risks, exposing them to security breaches and privacy concerns as potential attackers may successfully disclose personal information stored in vulnerable devices. For instance, in April 2019, around 40 million attacks were detected against various Ecuadorian companies \cite{comercio3}, and later the same year, this country experienced the biggest data theft in history when a database containing ID numbers and citizens' personal information was publicly disclosed \cite{planv4}. These issues may become more damaging with the advent of the Internet of Things (IoT), which enables users to connect different smart devices to the Internet, exposing private information in exchange for more affordable and better user experience. Although security and privacy requirements should be considered in the design of these devices; unfortunately, these are deployed  with inadequate, incomplete, or poorly designed features. E.g., security cameras connected to the Internet using weak authentication could be easily hacked to record actions of the inhabitants of a house. Conversely, traditional risk management approaches are a shortcoming for security and privacy analysts, making them more prone to jump into quick decisions for handling emerging threats. Then, it has been recently suggested to adopt proactive strategies for mitigating security and privacy threats \cite{caceres17}. Regarding security threats, the ISO/IEC 27005 standard compiles a set of best practices for risk management \cite{iso2}. Although, this standard does not address privacy issues directly, governmental initiatives such as the European General Data Protection Regulation (GDPR) \cite{gdpr11,yanez21} have emerged to fill this gap. As a result, whilst GDPR pinpoints privacy aspects as fundamental citizen rights, ISO 27005 may be used as a blueprint to implement them \cite{caceres17}, and possibly deploy security and privacy controls alike. Therefore, as suggested by \cite{everet5}, we aim to establish a synergy between security and privacy aspects by introducing a novel approach to risk management. Our proposal uses ISO 27005 for risk management as a baseline, using STRIDE as a methodology for threat identification \cite{scandari010}. Later, during risk analysis, inherent risk levels \cite{rodriguez6} are identified using attack trees. Lastly, for mitigation, our proposal identifies both ISO 27001/27002 security controls, and GDPR-based privacy controls in order to evaluate the corresponding residual risk levels using defense trees.

The rest of the paper continues as follows: In Sect. \ref{sec-rel-wrk}, related work is presented. In Sect. \ref{sec-bckg}, our research background is provided before introducing the proposed risk management approach in Sect. \ref{sec-frmwrk}. In Sect. \ref{sec-case-study}, an IoT-based computer architecture is used as a case study. Finally, results are discussed, and conclusions are given in Sect.~\ref{sec-conclude}.

\section{Related Work}
\label{sec-rel-wrk}
In this section, the most relevant contributions in the last 3 years are grouped, according to their characteristics, in 4 groups:

\begin{enumerate}
	\item \textit{Reference frameworks and ISO 27000 implementation guidelines}\cite{pino16, isorisk, isorisk2} which explain the correct deployment of ISO 27000. These works do not cover the protection of sensitive data.
	\item \textit{ISO 27000 Analysis and Risk Evaluation}\cite{isorisk, isorisk2} where extended policies are proposed based on this Standard. Although these studies are a great support for organizations, the lack of security controls to protect personal data is a limitation for risk evaluation in emerging computing architectures.	
	\item \textit{Literature reviews on ISO 27000}\cite{isogdrp2, caceres17, isogdrp} which cover a broad spectrum, from theoretical analysis to technical reviews, including some local regulations. Although these documents are excellent sources for explaining the application of this Standard in different regions and jurisdictions, its contribution is very limited. This is mainly due to poor threat mitigation strategies which overlook personal data protection.
	\item \textit{GDPR-compliant data protection considerations}\cite{isogdrp2, caceres17, isogdrp, yanez21} which propose a control sets for the proper handling of personal information. Nonetheless, these works do not to include a risk management cycle aligned with privacy controls.
\end{enumerate}

Overall, these works demonstrate that the ISO 27000 standard is important for securing the operation of computer systems. However, their lack of maturity is evident, as majority of them do not consider privacy aspects. Although few works have been focused on GDPR, they have not been tested in risk assessment scenarios, which make their applicability arguable. Hence, our contribution focuses exactly on this, providing a risk management approach that fills the gap between security and privacy controls. In the next section, we explain the background of our proposal before introducing our method in Sect. \ref{sec-frmwrk}. 

\section{Background}
\label{sec-bckg}
Securing information requires meeting acceptable levels of availability, integrity and confidentiality, which could be overlapped with privacy aspects such as anonymity and trust. Although these characteristics may or may not be enforced, for preventing security and privacy violations, quantifying threat materialization is required by means of an adequate risk management strategy\cite{magerit7}. First of all, computer systems, or \textit{assets} are \textit{threatened} by known \textit{vulnerabilities} (inherent weaknesses). A \textit{threat} to the security and privacy of a system is any (internal or external) action that may exploit a vulnerability in such system. Subsequently, \textit{risk} is the likelihood of \textit{threat materialization} on one or more assets. Thus, \textit{risk assessment} is an important information security process that quantifies the \textit{impact} of a threat, and its probability to be successful when vulnerable assets are exploited \cite{iso2}. Currently, either qualitative or quantitative risk assessment methods could be applied \cite{iso015}. However, these are usually reactive; i.e., applied to known threats, instead of being proactive, which requires \textit{threat identification and analysis} at early stages~\cite{owasp35}. Thus, Microsoft introduced \textit{STRIDE} \cite{scandari010}, as a threat modelling method for the identification and classification of threats within six categories: (i) Spoofing, (ii) Tampering, (iii) Repudiation, (iv) Information Disclosure, (v) Denial of Service and (vi) Elevation of Privilege. Similarly, this threat modelling technique can be complemented by using attack/defense trees \cite{kordy10, kordy29, sonderen33} as graphical tree-like representations of attack scenarios in which threats are seeing as tree nodes, and associated within attack paths (attack vectors) which may be executed to compromise a computer system. Finally, risk management requires \textit{identifying countermeasures} or controls. For this task, the ISO 27005 standard \cite{iso2} outlines a risk management life cycle in which security controls defined in the ISO 27001/27002 standards are used to mitigate potential threats. However, these control lack of privacy considerations, urging governmental agencies to propose alternative solutions. Hence, in 2016, the European Union issued the \textit{General Data Protection Regulation (GDPR)} in order to regulate personal information exchange as well as its storage and processing \cite{gdpr11, yanez21}. Our work integrates GDPR and ISO 27005 along with threat modelling practices into a risk management approach for handling security and privacy threats alike. This approach is explained in detail in the next section.


\section{Introducing a GDPR-compliant Threat-Model-Based Risk Management Approach}
\label{sec-frmwrk}

The ISO 27005 standard explains a process for \textit{risk management}, including assessment, treatment, communication, monitoring and acceptance \cite{agrawal27}.
Although our proposal uses ISO 27005 as a baseline, unlike its original counterpart, \textit{Risk Treatment} is divided into \textit{Control and Treatment}. In the next Sections, the phases in which our approach is divided into are explained in detail.

\subsection{Context Establishment}
This phase requires identifying the \textit{threat surface}, or an scenario, in which a computer system may be compromised. Such scenario is comprised of \textit{system and attack models}, which can be determined using a systematic literature search \cite{ramirez31} of reports about security breaches affecting sensitive personal data. Then, a \textit{system model} becomes a set of functional specifications and operation requirements of a vulnerable computer system. On the other hand, an \textit{attack model} represents security assumptions in which threats could compromise such a system. By using system and attack models, \textit{an initial threat catalogue} can be identified.

\subsection{Risk Assessment and Control}
\label{sec-rsk-assess}
\textit{Assessing risks} requires both threats and attacker profiles in order to define \textit{attack and defense trees} for quantifying both \textit{inherent and residual risk levels}.
Whilst an \textit{attack tree} is a decision tree structure that identifies plausible paths (or edges) for attackers to successfully execute threats in a vulnerable system \cite{sonderen33}, \textit{defense trees}, outlines controls that aim to counterattack such harmful events.

These trees are comprised of a \textit{root node} which, depicts the main attack goal, and several \textit{leaf or child nodes} \textit{grouped in [AND/OR] sub-goals}. For quantifying risks, node attributes in the tree are calculated using the equations defined in \cite{edge28, kordy29} as follows:
\begin{itemize}
	\item \textbf{Probability} is the likelihood of exploiting a $node_i$ or deploying a countermeasure, as defined in Eq. (\ref{eq-03}).
	\begin{equation}
	\label{eq-03}
	prob_i =
	\begin{cases}
	u & \text{if $prob_i < 0.05$}\\
	l & \text{if $0.05 \leq prob_i < 0.25$}\\
	m\prime & \text{if $0.25 \leq prob_i < 0.75$}\\
	h & \text{if $prob_i \geq 0.75$}\\
	c & \text{if $prob_i \text{ is close to } 1$}
	\end{cases}       
	\end{equation}
	
	The propagation of probability values upwards in the tree is defined in Eq. (\ref{eq-10}) and Eq. (\ref{eq-11}), where i corresponds to the number of converging nodes:
	\begin{itemize}
		\item In AND nodes:
		\begin{equation}
		\label{eq-10}
		prob_{and} = \prod_{i=1}^{n}prob_i
		\end{equation}
		\item In OR nodes:
		\begin{equation}
		\label{eq-11}
		prob_{or} = 1 - \prod_{i=1}^{n}(1 - prob_i)
		\end{equation}
	\end{itemize}
	
	Probability is similarly calculated in defense trees  as in Eq. (\ref{eq-15}), being i the number of converging nodes:
	
	\begin{equation}
	\label{eq-15}
	prob_{def} = prob_i \times \left(1 - \frac{prob_i}{cost_i}\right)
	\end{equation}
	
	\item \textbf{Cost} represents the amount of money required to exploit a $node_i$, or deploy countermeasures as in Eq. (\ref{eq-01}):
	\begin{equation}
	\label{eq-01}
	cost_i =
	\begin{cases}
	1 & \text{if cost is low}\\
	2 & \text{if cost is medium}\\
	3 & \text{if cost is high}
	\end{cases}       
	\end{equation}
	
	The propagation of cost upwards in the tree is defined in Eq.(\ref{eq-06}) and Eq. (\ref{eq-07}), where i corresponds to the number of converging nodes:
	\begin{itemize}
		\item In AND nodes:
		\begin{equation}
		\label{eq-06}
		cost_{and} = \sum_{i=1}^{n}cost_i
		\end{equation}
		\item In OR nodes:
		\begin{equation}
		\label{eq-07}
		cost_{or} = \frac{\sum_{i=1}^{n} prob_i \times cost_i}{\sum_{i=1}^{n} prob_i}
		\end{equation}
	\end{itemize}

	\item \textbf{Impact} is the severity of exploiting a $node_i$, as defined in Eq. (\ref{eq-02}), where a value of 10 denotes a system completely compromised, inoperable or destroyed. :
	\begin{equation}
	\label{eq-02}
	imp_i =
	\begin{cases}
	m & \text{if $1 \leq imp_i \leq 3$}\\
	n & \text{if $4 \leq imp_i \leq 6$}\\
	s & \text{if $7 \leq imp_i \leq 9$}\\
	10 & \text{otherwise}
	\end{cases}       
	\end{equation}
	Likewise, the impact propagation upwards in the tree is defined in Eq. (\ref{eq-08}) and Eq. (\ref{eq-09}), being i the number of converging nodes:
	\begin{itemize}
		\item In AND nodes:
		\begin{equation}
		\label{eq-08}
		imp_{and} = \frac{10^n - \prod_{i=1}^{n}(10 - imp_i)}{10^{(n-1)}}
		\end{equation}
		\item In OR nodes:
		\begin{equation}
		\label{eq-09}
		imp_{or} = \max^n_{i=1} (imp_i)
		\end{equation}
	\end{itemize}

	For the defense tree, impact is calculated as in Eq. (\ref{eq-14}), where i corresponds to the number of converging nodes,and effectiveness \textit{efc} $\in [0, 1]$:
	
	\begin{equation}
	\label{eq-14}
	imp_{def} = imp_i \times \frac{imp_i \times \textit{efc}_i}{cost_i \times 10}
	\end{equation}

	\item \textbf{Skill} is the required technical or social skill level for the attacker or defender to succeed, as defined in Eq. (\ref{eq-04}).
	\begin{equation}
	\label{eq-04}
	skill_i =
	\begin{cases}
	0.25 & \text{if skill low}\\
	0.5 & \text{if skill is medium}\\
	1 & \text{if skill is high}\\
	1.25 & \text{if skill is very high}
	\end{cases}       
	\end{equation}
	
	Skill values propagate upwards in the tree as defined in Eq. (\ref{eq-12}) and Eq. (\ref{eq-13}), being i the number of converging nodes:
	\begin{itemize}
		\item In AND nodes:
		\begin{equation}
		\label{eq-12}
		skill_{and} = \max^n_{i=1} (skill_i)		
		\end{equation}
		\item In OR nodes:
		\begin{equation}
		\label{eq-13}
		skill_{or} = \min^n_{i=1} (skill_i)
		\end{equation}
	\end{itemize}

	\item \textbf{Risk} estimates the degree of exposure to a threat ($node_i$), as in Eq. \ref{eq-05}, where i refers to the number of converging nodes:
	\begin{equation}
	\label{eq-05}
	risk_i = \frac{prob_i \times skill_i \times imp_i}{cost_i}  
	\end{equation}
\end{itemize}
Conversely, \textit{Risk Control} is focused on mitigating risks\cite{magerit7} by identifying \textit{security and privacy controls} based on ISO 27001/27002 and GDPR, respectively. This involves considering the likelihood (probability) of materialization, and the impact on the asset if an attack occurs. Table \ref{tab03} shows an example of \textit{probability countermeasures} integrating both standards.
\begin{table*}[h!]
	\caption{Samples of Probability Controls aligned with ISO 27001 and GDPR}
	\label{tab03}
	\centering
	\resizebox{\textwidth}{0.1\textheight}{
		\begin{tabular}{l|c|c|c|c|c|c|c}
			\multicolumn{1}{c|}{\textbf{Contermeasure}} &
			\multicolumn{1}{l|}{\textbf{Code}} &
			\textbf{\begin{tabular}[c]{@{}c@{}}ISO 27001\\ Section\end{tabular}} &
			\textbf{\begin{tabular}[c]{@{}c@{}}GDPR\\ Section\end{tabular}} &
			\textbf{Type} &
			\textbf{Value} &
			\textbf{Cost} &
			\textbf{\begin{tabular}[c]{@{}c@{}}Final\\ Value\end{tabular}} \\ \hline
			\begin{tabular}[c]{@{}l@{}}Establish standards for secure configuration,\\ development and updating of systems.\end{tabular} &
			C3 &
			\begin{tabular}[c]{@{}c@{}}14.1.3\\ 14.2.1\\ 14.2.2\end{tabular} &
			49 &
			Probability &
			0.80 &
			2 &
			0.40 \\ \hline
			\begin{tabular}[c]{@{}l@{}}Establish controls for protection \\ against malicious code.\end{tabular} &
			C4 &
			\begin{tabular}[c]{@{}c@{}}12.2.1\\ 12.6.2\end{tabular} &
			49 &
			Probability &
			0.80 &
			1 &
			0.80 \\ \hline
			Establish access control &
			C5 &
			\begin{tabular}[c]{@{}c@{}}9.1.1\\ 9.1.2\end{tabular} &
			\begin{tabular}[c]{@{}c@{}}39\\ 64\\ 83\end{tabular} &
			Probability &
			0.75 &
			1 &
			0.75 \\
		\end{tabular}	
	}
\end{table*}

The final value of each countermeasure is calculated as the ratio between its initial value and its cost. I.e., it is assumed that encouragement for its implementation depends on its cost, so if the cost increases, motivation for its deployment decreases.

\subsection{Risk Treatment}

It involves finding out their \textit{residual risk} after applying countermeasures. I.e., each important threat in the \textit{attack tree} has a countermeasure in the corresponding \textit{defense tree}. The latter uses the security-privacy controls devised in Table \ref{tab03} so that residual risk can be calculated.

\subsection{Risk Communication, Monitoring and Acceptance}

Communicating and validating a risk management strategy depends on \textit{analysing controls effectiveness}. In our proposal, this is done by comparing inherent and residual risk levels; i.e., before and after applying countermeasures.
As a final step, reviewing the outcomes of the \textit{effectiveness assessment} is important for monitoring risk levels and accepting them. The outcomes of this review will help organizations to decide whether it is necessary to start a new risk management cycle until risk levels are reduced, or accept the current levels. In the next section, an IoT network will be used as case study to apply our risk management proposal; particularly to demonstrate its applicability for risk treatment, communication, monitoring and acceptance.

\section{Case Study: Risk Assessment of an IoT Network}
\label{sec-case-study}
The Internet of Things (IoT) is a network of interconnected smart devices which, if misconfigured, may provide access to intimate aspects of the life of their users. E.g. security cameras that are recording the household. Besides its obvious benefits for surveillance and physical security, there are evident risks for user's privacy as their personal activities, and other information, could be exfiltrated to the Internet without consent. In the next sections, our proposed risk management approach is applied to an IoT scenario.

\subsection{Phase 1: Context Establishment}

Following the method proposed in \cite{ramirez31}, a systematic literature search was conducted over the Internet to discover threat reports related to security breaches affecting the Internet of Things during the year 2020. The main results are summarized in Table \ref{tab04} that shows that the main objective of IoT attackers is \textit{information~theft}.

\begin{table}[h!]
	\caption{Summary of Common IoT Security Breaches}
	\label{tab04}
	\centering
	\resizebox{\textwidth}{0.08\textheight}{
		\begin{tabular}{l|l|l}
			\multicolumn{1}{c|}{\textbf{Device}} &
			\multicolumn{1}{c|}{\textbf{\begin{tabular}[c]{@{}c@{}}Compromised Information\end{tabular}}} &
			\multicolumn{1}{c}{\textbf{\begin{tabular}[c]{@{}c@{}}Vulnerabilities Exploited\end{tabular}}} \\ \hline
			IP speakers &
			\begin{tabular}[c]{@{}l@{}}Audios of children and their surroundings.\end{tabular} &
			\begin{tabular}[c]{@{}l@{}}Private information exposed to the\\Internet through open ports.\end{tabular} \\ \hline
			Vehicles &
			\begin{tabular}[c]{@{}l@{}}Information such as address, geolocation\\and encrypted passwords.\end{tabular} &
			\begin{tabular}[c]{@{}l@{}} Lack of access control in devices installed\\in vehicles which stored important information.\end{tabular} \\ \hline
			Televisions &
			Images and audio. &
			\begin{tabular}[c]{@{}l@{}}Private information exposed to the Internet\\ through lack of access control on open ports.\end{tabular} \\ \hline
			\begin{tabular}[c]{@{}l@{}}Security cameras\end{tabular} &
			Security recordings &
			Default passwords not changed.
		\end{tabular}
	}
\end{table}

\subsection{Phase 2: Risk Assessment}

Based on the security breaches in Table \ref{tab04}, potential threats, and their risks, are analysed. The assumptions for this phase are explained as follows:

\subsubsection{System Model:}
In Fig. \ref{fig03-stride}, an IoT scenario is proposed to analyse possible threats that may affect an IoT infrastructure composed of these assets:

\begin{figure*}[h!]
	\centering
	\includegraphics[width=\textwidth, height=0.3\textheight]{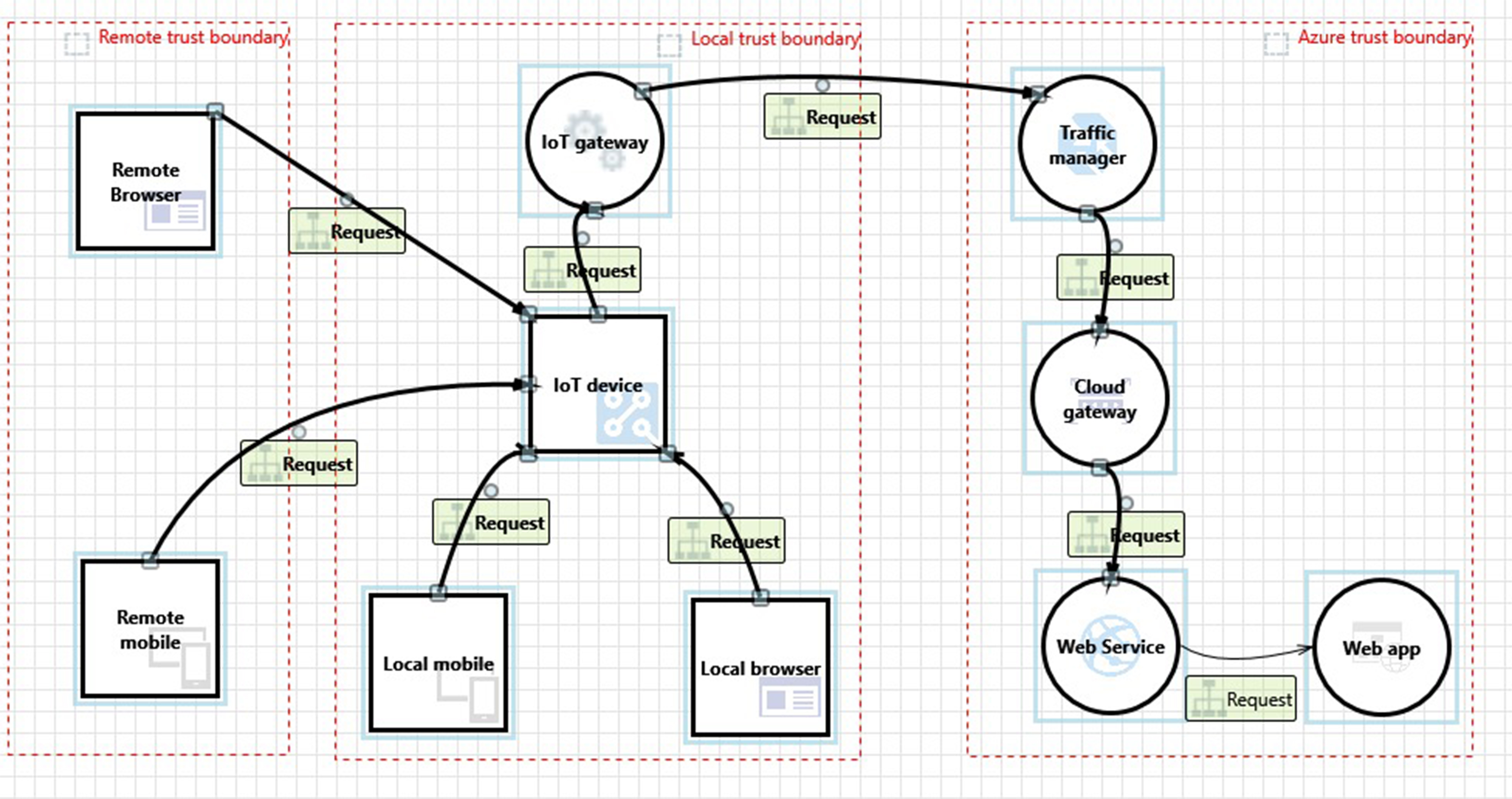}
	\caption{IoT-based Experimental Infrastructure}
	\label{fig03-stride}
\end{figure*}

\begin{itemize}
	\item IoT devices of any kind (monitors, vehicle controllers, etc.) are connected within a local trust boundary. These devices are assumed to operate with default credentials or with any. They are also connected to an IoT gateway that is in charge of redirecting traffic from the devices towards a traffic manager deployed within an Azure Cloud infrastructure.
	\item Devices in the local trust boundary interact with trusted local mobile devices. As they are trusted, they can connect to open, default or misconfigured ports.
	\item The Azure trust boundary represents an infrastructure of third-party applications with on-demand IoT services; such as storing surveillance information, geolocation metadata, etc.
	\item An external trust boundary hosts remote or external mobile devices which are less trusted, and used to interact with the IoT devices. E.g. mobile apps or web interfaces for remote operation of casting devices and security cameras.
\end{itemize}

\subsubsection{Attack Model:}
Considering the guidelines provided in \cite{peffers32}, it can be inferred that potential attackers could be external users of an IoT infrastructure who, without being the devices' owners, may access them in order to carry out malicious activities. E.g., credential misuse, physical port mishandling, etc.
Moreover, considering the potential breaches, the attacker profile, and the system model, any adversary will need an exposed port AND a lack of/default password to either ignore, hinder, misuse or tamper with a device\cite{iot}. 

\subsubsection{Initial Threat Catalogue:}
For analysing threats, Microsoft Threat Modelling Tool was used\footnote{MS Threat Modelling Tool available at https://bit.ly/3ussA00}, considering the assumptions described in both system and attack models. The resulting threat catalogue is shown in Table \ref{tab05}. For simplicity, only the highest-risk threats are considered, discarding those that do not affect privacy in order to model the attack/defense tree.
\begin{table}[h!]
	\caption{Initial Threat Catalogue with STRIDE Categories}
	\label{tab05}
	\resizebox{\textwidth}{0.24\textheight}{
		\begin{tabular}{@{}c|l|c|c|l@{}}
			\textbf{Id} &
			\multicolumn{1}{c|}{\textbf{Threat}} &
			\textbf{Asset} &
			\textbf{\begin{tabular}[c]{@{}c@{}}STRIDE\\ Category\end{tabular}} &
			\multicolumn{1}{c}{\textbf{Description}} \\
			\hline
			B1 &
			\begin{tabular}[c]{@{}l@{}}Eavesdropping the communication between\\ the device and the field gateway.\end{tabular} &
			Gateway &
			\begin{tabular}[c]{@{}c@{}}Information\\ Disclosure\end{tabular} &
			\begin{tabular}[c]{@{}l@{}}Private information exposed to the Internet\\ through open ports.\end{tabular} \\
			B2 &
			Gaining access to sensitive data from log files. &
			Gateway &
			\begin{tabular}[c]{@{}c@{}}Information\\ Disclosure\end{tabular} &
			\begin{tabular}[c]{@{}l@{}}Private information exposed to the Internet\\ through open ports.\end{tabular} \\
			B3 &
			\begin{tabular}[c]{@{}l@{}}Obtaining access to sensitive data by sniffing\\ traffic from Mobile client.\end{tabular} &
			Gateway &
			\begin{tabular}[c]{@{}c@{}}Information\\ Disclosure\end{tabular} &
			\begin{tabular}[c]{@{}l@{}}Private information exposed to the Internet\\ through open ports.\end{tabular} \\
			B4 &
			Reversing weakly encrypted or hashed content. &
			Device &
			\begin{tabular}[c]{@{}c@{}}Information\\ Disclosure\end{tabular} &
			\begin{tabular}[c]{@{}l@{}}Non-password protected interfaces, or\\ default password used.\end{tabular} \\
			B5 &
			Exploiting unused services in Gateway. &
			Gateway &
			\begin{tabular}[c]{@{}c@{}}Elevation of\\ Privileges\end{tabular} &
			\begin{tabular}[c]{@{}l@{}}Private information exposed to the Internet\\ through open ports.\end{tabular} \\
			B6 &
			\begin{tabular}[c]{@{}l@{}}Unauthorized access to privileged features\\ on Device.\end{tabular} &
			Device &
			\begin{tabular}[c]{@{}c@{}}Elevation of\\ Privileges\end{tabular} &
			\begin{tabular}[c]{@{}l@{}}Remote control of devices installed on\\ vehicles.\end{tabular} \\
			B7 &
			\begin{tabular}[c]{@{}l@{}}Unauthorized access to privileged features\\ on Gateway.\end{tabular} &
			Gateway &
			\begin{tabular}[c]{@{}c@{}}Elevation of\\ Privileges\end{tabular} &
			\begin{tabular}[c]{@{}l@{}}Private information exposed to the Internet\\ through open ports.\end{tabular} \\
			B8 &
			Executing unknown code on device. &
			Device &
			Tampering &
			The device allowed to execute external code. \\
			B11 &
			\begin{tabular}[c]{@{}l@{}}Tampering with devices and extract \\ cryptographic key material from it.\end{tabular} &
			Device &
			Tampering &
			\begin{tabular}[c]{@{}l@{}}Non-password protected interfaces, or\\ default password used.\end{tabular} \\
			B13 &
			\begin{tabular}[c]{@{}l@{}}Unauthorized access to IoT Field Gateway\\ to tamper with its OS.\end{tabular} &
			Gateway &
			Tampering &
			\begin{tabular}[c]{@{}l@{}}Private information exposed to the Internet\\ through open ports.\end{tabular} \\
			B14 &
			Spoofing a device to connect to field gateway. &
			Gateway &
			Spoofing &
			\begin{tabular}[c]{@{}l@{}}Private information exposed to the Internet\\ through open ports.\end{tabular} \\
			B15 &
			\begin{tabular}[c]{@{}l@{}}Gaining access to the field gateway by using \\ default login credentials.\end{tabular} &
			Gateway &
			Spoofing &
			\begin{tabular}[c]{@{}l@{}}Private information exposed to the Internet\\ through open ports.\end{tabular} \\
			B16 &
			Reusing authentication tokens of Device &
			Device &
			Spoofing &
			\begin{tabular}[c]{@{}l@{}}Non-password protected interfaces, or\\ default password used.\end{tabular}
		\end{tabular}
	}
\end{table}
\subsubsection{Inherent Risk Assessment:}

For assessing inherent risks, we use both the threat catalogue and the attacker profile to create attack trees in ADTool\footnote{ADTool available at https://bit.ly/3xRetDu}.

The notation of attack and defense trees is shown in Table \ref{tab02}. 
\begin{table}[h!]
	\caption{Attack/Defense Tree Representation}
	\label{tab02}
	\centering
	\resizebox{0.6\textwidth}{0.06\textheight}{
		\begin{tabular}{c|c|c|c}
			\begin{tabular}[c]{@{}c@{}}Root Node\end{tabular} &
			Countermeasure &
			\begin{tabular}[c]{@{}c@{}}OR Branch\end{tabular} &
			\begin{tabular}[c]{@{}c@{}}AND Branch\end{tabular}\\
			\midrule
			\includegraphics[width=1.4cm, height=0.70cm]{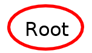}&
			\includegraphics[width=1cm, height=0.6cm]{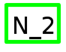} &
			\includegraphics[width=1.6cm, height=1.6cm]{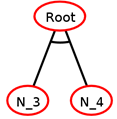} &
			\includegraphics[width=1.6cm, height=1.6cm]{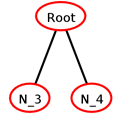}
		\end{tabular}
	}
\end{table}
It is assumed that an attacker will attempt to steal private information by compromising either the IoT gateway or the device itself. For better understanding, the attack goal is split in two subtrees per each attack objective. One of these attack trees is shown in Fig. \ref{fig06-attacks}. Next, metrics of probability, cost, impact and skill are used for inherent risk assessment as follows:


\begin{figure}[h!]
	\centering
		\includegraphics[clip,width=\textwidth, height=0.33\textheight]{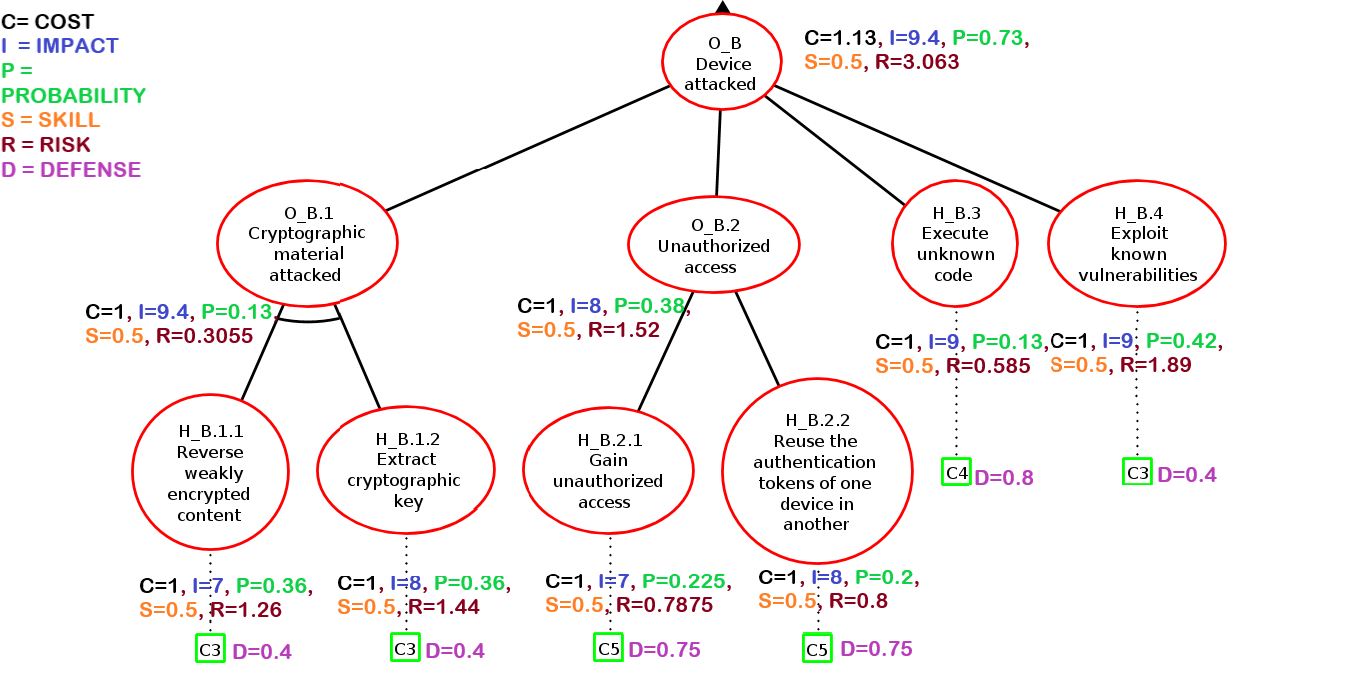}
		\caption{Attack/Defense Tree for IoT devices with its corresponding metrics.}
	\label{fig06-attacks}
\end{figure}
\begin{itemize}
	\item For leaf nodes denoted by \textit{H}, these values are estimated within discrete intervals defined in Eqs. (\ref{eq-03}), (\ref{eq-01}), (\ref{eq-02}) and (\ref{eq-04}).
	\item For each AND/OR node denoted by \textit{O}, values of probability, cost, impact and skill are propagated upwards, using their corresponding equations, as defined in Sect. \ref{sec-rsk-assess}.
	\item Finally, inherent risk is calculated using Eq. (\ref{eq-05}).
\end{itemize}
For instance, the attribute values in $O\_B.2$ (OR Node) are an aggregation of the values in $H\_B.2.1$ and $H\_B.2.2$, which can be calculated using Eqs. (\ref{eq-07}) for cost, (\ref{eq-09}) for impact, (\ref{eq-13}) for skill, (\ref{eq-11}) for probability, and (\ref{eq-05}) for risk:
\begin{equation*} 
\begin{split}
cost_{O\_B.2}&		= \frac{(0.225 \cdot 1)+(0.20 \cdot 1)}{(0.225+ 0.20)} = \frac{(0.225+0.20)}{0.425} = 1 \\
imp_{O\_B.2}&		=  \max{(7 , 8)} = 8\\
skill_{O\_B.2}&	=  \min{(0.5, 0.5)} =  0.5\\
prob_{O\_B.2}&	= 1-((1-0.225) \times (1-0.20)) = 0.38\\
risk_{O\_B.2}&		=  \frac{(0.38 \times 8 \times 0.5)}{1} = 1.52
\end{split}
\end{equation*}

\subsection{Phase 3: Risk Treatment}
A fundamental stage of our proposal is risk mitigation, using security-privacy controls as defined in Table \ref{tab03}. 
As it can be seen in Fig. \ref{fig06-attacks}, countermeasures have been applied to leaf nodes so that any action taken to mitigate them will aggregate controls upwards to minimize adverse consequences in AND/OR nodes. In defense trees, such countermeasures mitigate either the probability or the impact of an attack; the former to reduce chances of materialization, and the latter to limit the potential damage. Hence, after applying countermeasures, the residual risk can be calculated as follows:
\begin{itemize}
	\item For leaf nodes denoted by \textit{H}:
		\begin{itemize}
			\item Cost and skill values are estimated within discrete intervals defined in Eqs. (\ref{eq-01}) and (\ref{eq-04}).
			\item As only probability countermeasures are used, these values are calculated using Eq. (\ref{eq-15}). If impact countermeasures were considered, Eq. (\ref{eq-14}) should be also applied.
		\end{itemize}
	\item Like in attack trees, Eqs. (\ref{eq-15}) and (\ref{eq-14}) are used to propagate probability and impact values upwards in AND/OR nodes denoted by \textit{O}.  
	\item Similarly, residual risk is calculated using Eq. (\ref{eq-05}).
\end{itemize}
\begin{table*}[h!]
	\caption{Resulting Risk Analysis and Treatment Metrics: A Comparison of Inherent and Residual Risk Levels}
	\label{tab06}
	\centering
	\resizebox{\textwidth}{0.23\textheight}{
		\begin{tabular}{c|c|c|c|c|c|c|c|c|c|c|c|c|c|c}
			\multicolumn{7}{c|}{\textbf{Risk Assessment}} &
			\multicolumn{7}{c|}{\textbf{Risk Treatment}} &
			\multirow{2}{*}{\textbf{\begin{tabular}[c]{@{}c@{}}\% \\ Reduction\end{tabular}}} \\
			\cline{1-14}
			\textbf{\begin{tabular}[c]{@{}c@{}}Threat\\ Code\end{tabular}} &
			\textbf{Node} &
			\textbf{Cost} &
			\textbf{Impact} &
			\textbf{Skill} &
			\textbf{Probability} &
			\textbf{\begin{tabular}[c]{@{}c@{}}Inherent\\ Risk\end{tabular}} &
			\textbf{\begin{tabular}[c]{@{}c@{}}Control \\ Code\end{tabular}} &
			\textbf{Value} &
			\textbf{Cost} &
			\textbf{Impact} &
			\textbf{Skill} &
			\textbf{Probability} &
			\textbf{\begin{tabular}[c]{@{}c@{}}Residual\\ Risk\end{tabular}} &
			\\ \hline
			B1  & H\_A.1.1 & 1    & 7   & 0.5 & 0.7  & 2.45       & C3 & 0.4  & 1    & 7   & 0.5 & 0.42  & 1.47    & 40.0\% \\
			B3  & H\_A.1.2 & 1    & 7   & 0.5 & 0.65 & 2.275      & C3 & 0.4  & 1    & 7   & 0.5 & 0.39  & 1.365   & 40.0\% \\
			B5  & H\_A.2   & 1    & 8   & 0.5 & 0.7  & 2.8        & C3 & 0.4  & 1    & 8   & 0.5 & 0.42  & 1.68    & 40.0\% \\
			B9  & H\_A.3   & 1    & 9   & 0.5 & 0.65 & 2.925      & C4 & 0.8  & 1    & 9   & 0.5 & 0.13  & 0.585   & 80.0\% \\
			B2  & H\_A.4.1 & 1    & 8   & 0.5 & 0.7  & 2.8        & C5 & 0.75 & 1    & 8   & 0.5 & 0.175 & 0.7     & 75.0\% \\
			B7  & H\_A.4.2 & 1    & 8   & 0.5 & 0.9  & 3.6        & C5 & 0.75 & 1    & 8   & 0.5 & 0.225 & 0.9     & 75.0\% \\
			B12 & H\_A.4.3 & 1    & 8   & 0.5 & 0.5  & 2          & C3 & 0.4  & 1    & 8   & 0.5 & 0.3   & 1.2     & 40.0\% \\
			B15 & H\_A.4.4 & 1    & 8   & 0.5 & 0.9  & 3.6        & C3 & 0.4  & 1    & 8   & 0.5 & 0.54  & 2.16    & 40.0\% \\
			B14 & H\_A.4.5 & 2    & 9   & 0.5 & 0.6  & 1.35       & C3 & 0.4  & 2    & 9   & 0.5 & 0.48  & 1.08    & 20.0\% \\
			B4  & H\_B.1.1 & 1    & 7   & 0.5 & 0.6  & 2.1        & C3 & 0.4  & 1    & 7   & 0.5 & 0.36  & 1.26    & 40.0\% \\
			B11 & H\_B.1.2 & 1    & 8   & 0.5 & 0.6  & 2.4        & C3 & 0.4  & 1    & 8   & 0.5 & 0.36  & 1.44    & 40.0\% \\
			B6  & H\_B.2.1 & 1    & 7   & 0.5 & 0.9  & 3.15       & C5 & 0.75 & 1    & 7   & 0.5 & 0.225 & 0.7875  & 75.0\% \\
			B8  & H\_B.2.2 & 1    & 8   & 0.5 & 0.8  & 3.2        & C5 & 0.75 & 1    & 8   & 0.5 & 0.2   & 0.8     & 75.0\% \\
			B10 & H\_B.3   & 1    & 9   & 0.5 & 0.65 & 2.925      & C4 & 0.8  & 1    & 9   & 0.5 & 0.13  & 0.585   & 80.0\% \\
			-   & H\_B.4   & 1    & 9   & 0.5 & 0.7  & 3.15       & C3 & 0.4  & 1    & 9   & 0.5 & 0.42  & 1.89    & 40.0\% \\
			B13 & O\_A.1   & 2    & 9.1 & 0.5 & 0.46 & 1.0465     & -  & -    & 2    & 9.1 & 0.5 & 0.16  & 0.364   & 65.2\% \\
			-   & O\_A.4   & 1.17 & 9   & 0.5 & 1    & 3.84615385 & -  & -    & 1.23 & 9   & 0.5 & 0.87  & 3.18293 & 17.2\% \\
			-   & O\_B.1   & 2    & 9.4 & 0.5 & 0.36 & 0.846      & -  & -    & 2    & 9.4 & 0.5 & 0.13  & 0.3055  & 63.9\% \\
			-   & O\_B.2   & 1    & 8   & 0.5 & 0.98 & 3.92       & -  & -    & 1    & 8   & 0.5 & 0.38  & 1.52    & 61.2\% \\
			-   & O\_A     & 1.22 & 9.1 & 0.5 & 1    & 3.7295082  & -  & -    & 1.23 & 9.1 & 0.5 & 0.94  & 3.47724 & 6.8\%  \\
			-   & O\_B     & 1.13 & 9.4 & 0.5 & 1    & 4.15929204 & -  & -    & 1.12 & 9.4 & 0.5 & 0.73  & 3.06339 & 26.3\% \\
			-   & O\_T     & 1.18 & 9.4 & 0.5 & 1    & 3.98305085 & -  & -    & 1.18 & 9.4 & 0.5 & 0.98  & 3.90339 & 2.0\% 
	\end{tabular}}
\end{table*}
\subsection{Phase 4: Risk Communication, Monitoring and Acceptance}

Communicating risks requires analysing the effectiveness of the adopted strategy. Thus, Table \ref{tab06} shows results of inherent and residual risk calculated using attack and defense trees, respectively. Here, the values of residual risk are obtained after security-privacy controls have been used as countermeasures. As a consequence, sensitive information can be protected, reducing the risk in leaf nodes as well as AND/OR nodes to acceptable levels. Based on these results, we have demonstrated that:
\begin{itemize}
	\item The inherent risk generated by attacks to steal sensitive personal information has a very high impact, affecting the privacy of IoT users considerably.
	\item The proposed security-privacy controls for safeguarding users' privacy in IoT environments can significantly contribute to reduce the probability of attacks on leaf nodes.
\end{itemize}

However, despite having a risk reduction to levels up to 80 percent, a value of 2 per cent of reduction in the main attack goal ($O_T$-node) indicates the existence of a persistent threat that could not be fully mitigated.
In contrast, comparing the values between \textit{inherent and residual risk levels}, it can be said that the implementation of security controls based on GDPR and ISO 27001/27002 need to be strictly controlled as information theft is a persistent threat to data security and privacy. Henceforth, the proposed risk management cycle should be performed regularly to ensure its effectiveness; particularly, when IoT devices and users are interacting within different trust levels.

\section{Conclusion and Future Work}
\label{sec-conclude}
This research has introduced a novel threat-modelling method for risk assessment which, unlike previous work, integrates security and privacy controls into a simple, yet effective method, fully compliant with other ISO 27000-based frameworks such as Magerit \cite{magerit7}.
Our proposal utilizes both STRIDE and attack/defense trees, aligning these practices within the ISO 27005 risk management cycle. Whilst attack trees aid the identification and mitigation of threats, defense trees allow applying countermeasures; promoting the proactive identification of inherent and residual risks. Finally, even though our approach could be applied to any scenario, we demonstrated its resilience for analysing IoT-related threats.
\begin{figure*}[h!]
	\centering
	\begin{subfigure}[b]{0.48\textwidth}
		\includegraphics[clip,width=0.91\textwidth, height=0.28\textheight]{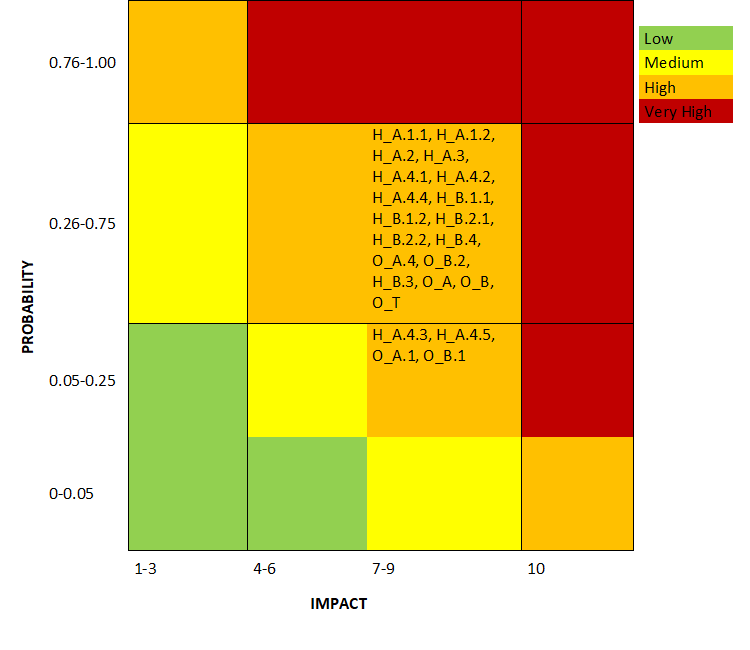}
		\caption{Inherent Risk Assessment}
		\label{fig07-inh-rsk}
	\end{subfigure}
	~
	\begin{subfigure}[b]{0.48\textwidth}
		\includegraphics[clip,width=0.9\textwidth, height=0.28\textheight]{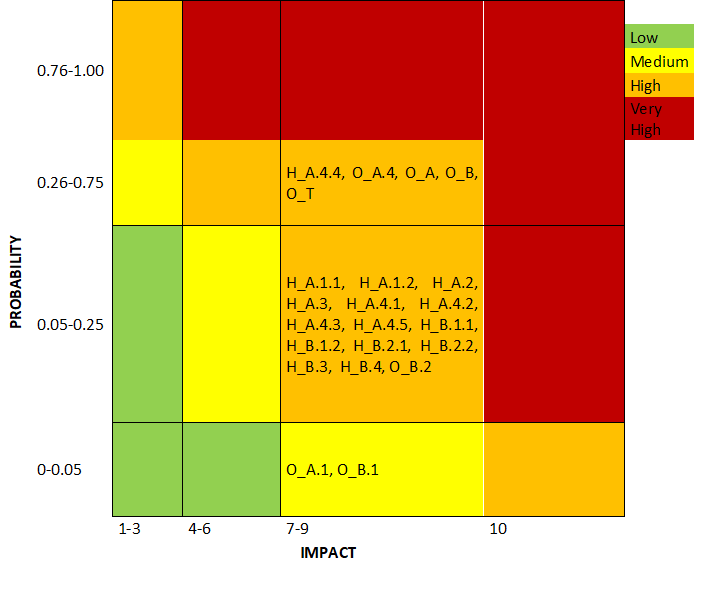}
		\caption{Residual Risk Assessment}
		\label{fig07-res-rsk}
	\end{subfigure}		
	\caption{Risk Level Variation before and after Applying Countermeasures.}
	\label{fig07-maps}
\end{figure*}

This becomes evident in Figure \ref{fig07-maps}, where the identified security-privacy controls in Table \ref{tab03} seem good enough to mitigate the chances of materialization of particular threats to users' privacy. However, although inherent and residual risk reduction is quite evident, it is not complete. This suggests the existence of persistent threats in IoT scenarios which requires in-depth analysis of more complex architectures. Despite its limitations, our contribution is a step forward towards integrating security and privacy which are important characteristics for the design and deployment of computer systems. In fact, as shown in our work, if GDPR \cite{gdpr11, yanez21} is used for designing countermeasures to prevent privacy violations, the adoption of one-sided security strategies can be avoided, even before any computer system is deployed in a production environment.


%
%
\bibliographystyle{./styles/bibtex/spphys}
\bibliography{./bibliography/biblios}

\begin{thebibliography}{10}
\providecommand{\url}[1]{{#1}}
\providecommand{\urlprefix}{URL }
\expandafter\ifx\csname urlstyle\endcsname\relax
  \providecommand{\doi}[1]{DOI \discretionary{}{}{}#1}\else
  \providecommand{\doi}{DOI \discretionary{}{}{}\begingroup
  \urlstyle{rm}\Url}\fi

\bibitem{iso2}
{ISO, International Organization for Standardization}.
\newblock {ISO/EC27000:2018} (2018).
\newblock \urlprefix\url{https://bit.ly/2PG7xaW}.
\newblock [Accessed 20 March 2021]

\bibitem{comercio3}
{El Comercio}.
\newblock Hackers launched a global offensive to attack state websites (2019).
\newblock \urlprefix\url{https://bit.ly/3di1Vvz}.
\newblock [Accessed 15 December 2020][From Spanish]

\bibitem{planv4}
{Plan V}.
\newblock {The worst data breach in the history of Ecuador exposed} (2019).
\newblock \urlprefix\url{https://bit.ly/3rwxIhm}.
\newblock [Accessed 15 December 2020][From Spanish]

\bibitem{caceres17}
I.M. {Lopes}, T.~{Guarda}, P.~{Oliveira}, in \emph{2019 14th Iberian Conference
  on Information Systems and Technologies (CISTI)} (2019), pp. 1--6.
\newblock \doi{10.23919/CISTI.2019.8760937}

\bibitem{gdpr11}
{European Parliament}, \emph{Regulation (EU) 2016/679} (Official Journal of the
  European Union, 2016).
\newblock \urlprefix\url{https://bit.ly/39rEVZQ}.
\newblock [Accessed 20 March 2021]

\bibitem{yanez21}
{European Union}, \emph{{Guidance: Security Measures for Personal Data
  Processing Article 22 of Regulation 45/2001}} (European data protection
  Supervisor, 2016).
\newblock \urlprefix\url{https://bit.ly/3czNfsA}.
\newblock [Accessed 20 March 2021]

\bibitem{everet5}
C.~Everett, Computer Fraud and Security \textbf{2011}(2), 5 (2011)

\bibitem{scandari010}
R.~Scandariato, K.~Wuyts, W.~Joosen, Springer-Verlag \textbf{2013}(1), 18
  (2013).
\newblock \urlprefix\url{https://doi.org/10.1007/s00766-013-0195-2}

\bibitem{rodriguez6}
F.~Roth, \emph{Visualizing Risk: The Use of Graphical Elements in Risk Analysis
  and Communications}.
\newblock 3RG report (Eidgen{\"o}ssische Technische Hochschule Z{\"u}rich,
  Center for Security Studies CSS, 2012).
\newblock \urlprefix\url{https://bit.ly/3rABw15}.
\newblock [Accessed 20 March 2021]

\bibitem{isogdrp2}
M.~Gaspar, S.~Popescu, Acta Technica Napocensis - Series: Applied Mathematics,
  Mechanics and Engineering \textbf{61}(3Spe), 85 (2018).
\newblock \urlprefix\url{https://bit.ly/3wcZVgO}.
\newblock [Accessed 25 March 2021]

\bibitem{pino16}
D.~Tjirare, {Designing a National Adoption Policy Framework for ISO/IEC 27000
  Standards Implementation in Namibia}.
\newblock Master's thesis, Namibia University of Science and Technology (2018).
\newblock \urlprefix\url{https://bit.ly/3doz1K2}.
\newblock [Accessed 28 March 2021]

\bibitem{isogdrp}
J.~Wichmann, K.~Sandkuhl, N.~Shilov, A.~Smirnov, F.~Timm, M.~Wißotzki, RTU
  Press \textbf{2020}(24), 31 (2020)

\bibitem{isorisk}
A.I.H. bin Suhaimi, N.~Noordin, M.F. bin Ya'kub, Journal of Physics: Conference
  Series \textbf{1551}, 012003 (2020).
\newblock \urlprefix\url{https://doi.org/10.1088/1742-6596/1551/1/012003}

\bibitem{isorisk2}
S.~Putra, M.~Gunawan, A.~Sobri, J.~Muslimin, Amilin, D.~Saepudin, 2020 8th
  International Conference on Cyber and IT Service Management (CITSM)
  \textbf{2020}(8), 1 (2020)

\bibitem{magerit7}
{Directorate General for Administrative Modernisation, Procedures and Promotion
  of e-Government}, \emph{Methodology for Information Systems Risk Analysis and
  Management} (Ministry of Finance and Public Administration, Spain, 2014)

\bibitem{iso015}
{International Organization for Standardization (ISO) \& International
  Electrotechnical Commission (IEC)}, \emph{{NTE INEN-ISO/IEC} 27005:2015
  Tecnolog\'ia de la Informaci\'on - T\'ecnicas de Seguridad - Gesti\'on de
  Riesgos en la Seguridad de la Informaci\'on} (2015).
\newblock [Spanish version]

\bibitem{owasp35}
OWASP.
\newblock {OWASP SAMM} (2020).
\newblock \urlprefix\url{https://owasp.org/www-project-samm/}.
\newblock [Accessed 15 December 2020]

\bibitem{kordy10}
B.~Kordy, P.~Kordy, S.~Mauw, P.~Schweitzer, CoRR \textbf{abs/1305.6829} (2013).
\newblock \urlprefix\url{http://arxiv.org/abs/1305.6829}.
\newblock [Accessed 20 March 2021]

\bibitem{kordy29}
A.~Bagnato, B.~Kordy, P.H. Meland, P.~Schweitzer, International Journal of
  Secure Software Engineering (IJSSE) \textbf{3}, 1 (2012)

\bibitem{sonderen33}
T.~Sonderen, {A Manual for Attack Trees}.
\newblock Master's thesis, University of Twente (2019).
\newblock \urlprefix\url{http://essay.utwente.nl/79133/}.
\newblock [Accessed 28 March 2021]

\bibitem{agrawal27}
V.~Agrawal, IEEE 4th International Conference on Cyber Security and Cloud
  Computing \textbf{2017}(1), 264 (2017)

\bibitem{ramirez31}
Y.~Xiao, M.~Watson, Journal of Planning Education and Research \textbf{39}(1),
  93 (2019).
\newblock \urlprefix\url{https://doi.org/10.1177/0739456X17723971}

\bibitem{edge28}
K.~Edge, G.~Dalton, R.~Raines, R.~Mills, IEEE Mil. Commun. \textbf{2007}(1), 1
  (2007)

\bibitem{peffers32}
K.~Peffers, T.~Tuunanen, M.A. Rothenberger, S.~Chatterjee, Journal of
  Management Information Systems \textbf{24}(3), 45 (2007).
\newblock \urlprefix\url{https://doi.org/10.2753/MIS0742-1222240302}

\bibitem{iot}
F.~Meneghello, M.~Calore, D.~Zucchetto, M.~Polese, A.~Zanella, 2020 8th
  International Conference on Cyber and IT Service Management (CITSM)
  \textbf{2020}(8), 1 (2020)

\end{thebibliography}
\end{document}